\documentclass[a4paper,12pt]{article}
\usepackage{epsfig}

\begin{document}

\def\beq{\begin{equation}}
\def\eeq{\end{equation}}
\def\bea{\begin{eqnarray}}
\def\eea{\end{eqnarray}}
\def\Dss{\frac{1}{3 s^2}}
\def\Dtt{\frac{1}{t^2}}
\def\Duu{\frac{1}{u^2}}
\def\Dtu{\frac{1}{t\,u}}
\def\Dts{\frac{1}{t\,s}}
\def\Dt{\frac{1}{t}}
\def\Du{\frac{1}{u}}
\def\Ds{\frac{1}{s}}
\newcommand{\dedouble}{ \stackrel{ \leftrightarrow }{ \partial } }
\newcommand{\deR}{ \stackrel{ \rightarrow }{ \partial } }
\newcommand{\deL}{ \stackrel{ \leftarrow }{ \partial } }
\def\dofig#1#2{\epsfxsize=#1\centerline{\epsfig{file=#2, width=14cm, 
height=5cm, angle=0}}}
\def\dofigs2#1#2#3{\centerline{\epsfxsize=#1\epsfig{file=#2, width=7.5cm, 
height=7.5cm, angle=-90}
\hfil\epsfxsize=#1\epsfig{file=#3,  width=7.5cm, height=7.5cm, angle=-90}}}
\def\makefigs4#1#2#3#4#5{
\centerline{\epsfxsize=#1\epsfig{file=#2, width=7.5cm, 
height=7.5cm, angle=-90}
\hfil\epsfxsize=#1\epsfig{file=#3,  width=7.5cm, height=7.5cm, angle=-90}}
\centerline{\epsfxsize=#1\epsfig{file=#4, width=7.5cm, 
height=7.5cm, angle=-90}
\hfil\epsfxsize=#1\epsfig{file=#5,  width=7.5cm, height=7.5cm, angle=-90}}
}
\renewcommand{\thefootnote}{\fnsymbol{footnote}}
\rightline{} \rightline{HIP-2002-42/TH}

\rightline{ROME1-1345/2002} 
\vspace{.3cm} 
{\Large
\begin{center}
{\bf Virtual graviton exchanges at the Z pole\\ in
large extra dimensions}

\end{center}}
\vspace{.3cm}

\begin{center}
Anindya Datta$^{1}$, Emidio Gabrielli$^{1}$, and Barbara Mele$^{2}$ \\
\vspace{.3cm}
$^1$\emph{
Helsinki Institute of Physics,
     POB 64, University of Helsinki, FIN 00014, Finland}
\\
$^2$\emph{Istituto Nazionale di Fisica Nucleare, Sezione di Roma,
and Dip. di Fisica, Universit\`a La Sapienza,
P.le A. Moro 2, I-00185 Rome, Italy}
\end{center}

\vspace{.3cm}
\hrule \vskip 0.3cm
\begin{center}
\small{\bf Abstract}\\[3mm]
\end{center}

In the framework of quantum gravity propagating in large extra
dimensions, the effects of virtual Kaluza-Klein gravitons on the
imaginary part of the amplitude of the process $e^+e^- \to f \bar{f}$
are analyzed at the $Z$ pole.  Notably, the interference of the
almost-continuous spectrum of the KK gravitons with the standard model
resonant amplitude is finite and predictable in terms of the
fundamental D-dimensional Plank scale $M_D$.  We find that, while the
virtual-graviton effect on total cross sections vanishes at
tree-level, both angular and polarization asymmetries are modified by
terms whose {\it relative} effect is at most of order $10^{-4}$ for
$M_D> 1$ TeV.  Possible shifts in $M_Z$, due to the 
changes in Breit-Wigner line shape are also discussed.

\begin{minipage}[h]{14.0cm}
\end{minipage}
\vskip 0.3cm \hrule \vskip 0.5cm
\section{Introduction}
As shown by Arkani-Hamed, Dimopoulos, and Dvali (ADD) a few years ago
\cite{ADD},
the existence of large extra dimensions with only gravity propagating in
the bulk,
provides an interesting alternative solution to the hierarchy 
problem.
A dramatic consequence of this scenario is that gravity 
might become strong at the TeV scale, hence making plausible
the possibility of detecting quantum gravity effects
at future collider experiments \cite{ED_string,GRW,PHEN_ED}.

In the ADD scenario, the Newton's constant
$G_N$ in the 3+1 dimensional space is related to  the corresponding Planck
scale $M_D$ in the $D=4+\delta$ dimensional space by \cite{ADD}
\beq
G_N^{-1}= 8 \pi R^{\delta} M_D^{2+\delta}
\label{newton}
\eeq
where $R$ is the radius of the compact manifold
assumed to be on a torus.
Therefore, the weakness of gravity might be due to the large size of the
compactified extra dimensional space.
In particular, if $M_D\sim $ TeV then
deviations from the Newton's law are expected
at distances of order $R < 10^{32/\delta -19}$ meters \cite{Gravity_test}.
This is compatible with the present experimental sensitivity in gravity tests,
which is above the millimeter scale. Then, assuming $M_D \sim $ TeV, 
Eq.(\ref{newton}) implies $\delta \ge 2$.

The effective Einstein theory in 3+1 dimension is then obtained by
integrating out compact extra dimensions. It 
describes massive Kaluza-Klein (KK) excitations
of the standard graviton field.\footnote{
For a detailed discussion about the
effective four-dimensional theory, see  \cite{GRW}.}
The KK excitations
are very narrowly spaced in comparison to the $M_D$ scale,
and can be often treated as a continuum.
Remarkably,
in the inclusive production (or virtual exchange) 
of KK gravitons, the sum over the allowed tower 
of KK states replaces the suppression factor $(E/M_P)^2$ 
associated to a single graviton production 
by the quantity $(E/M_D)^{2+\delta}$,
where $E$ is the typical energy of the process,
and $M_P$ is the Plank mass. 
Therefore, if $M_D$ is of order of TeV, quantum gravity effects may become 
accessible at future collider experiments.

In the framework of the ADD scenario,
two classes of processes have been analyzed
in $e^+e^-$ and hadron collisions \cite{ED_string,GRW,PHEN_ED,land} :
the direct production of  KK gravitons, 
detected as missing energy in the final states, 
and the virtual KK graviton exchange. 
In the latter case, the graviton exchange  induces local 
dimension-eight operators 
(associated with the square of the energy momentum-tensor)
that will affect, for instance, the standard four fermion processes. 
The real part of this kind of
contribution is divergent \cite{GRW}. The divergences can be regularized 
by introducing an unknown ultraviolet cutoff,  therefore 
loosing the predictivity in terms of the fundamental Plank scale $M_D$.
Indeed, in the effective Einstein 
theory in 3+1 dimension, arising by integrating out extra dimensions,
divergences can appear even at tree-level after summing over the KK
states of the graviton propagator. 
In the case of virtual graviton exchange,  in  many analyses 
the assumption is made
that the {\it divergent} real part of the amplitude ${\cal A}$
 is dominated 
by the lowest dimensional operator (of dimension eight in the case
considered)
\beq
Re\left[{\cal A}\right] = \frac{4\pi}{\Lambda_T^4} \left(T_{\mu \nu} T^{\mu\nu} -
\frac{1}{\delta +2} T_{\mu}^{\mu} T_{\nu}^{\nu}\right)
\label{effective}
\eeq
where $T_{\mu \nu}$ is the energy-momentum tensor. The new scale
$\Lambda_T$ is an effective cutoff, assumed to be 
of order of $M_D$, introduced to parametrize unknown new physics 
contributions in the ultraviolet region.

On the other hand, the imaginary part $Im[{\cal A}]$ is finite, 
being connected to the branch-cut singularity of
real graviton emission \cite{GRW}.

Up to now, phenomenological analyses 
have considered virtual KK graviton contributions to the standard model (SM) 
processes in kinematical regions far from the resonances 
of the SM spectrum.
In this case, the leading contribution to the physical process is due 
to the interference of the amplitude $Re[{\cal A}]$ 
with the real part of the corresponding SM amplitude, 
and is suppressed by $(E/\Lambda_T)^4$, where $E$ 
(assumed to be $E< \Lambda_T$) is the typical energy of the process.

On the other hand, when considering SM resonant processes 
interfering with virtual KK graviton graphs, the
corresponding interference is dominated by $Im[{\cal A}]$.
Hence, it is finite, and predictable in terms  of the fundamental 
Plank scale $M_D$.

In this paper we study an example of the latter class of processes,
that is the two-fermion production in $e^+e^-$ collisions, 
$e^+e^-\to f\bar{f}$, 
in the resonant region of the Z pole. 
We compute the {\it finite} interference of the SM amplitude
with the $e^+e^-\to G \to f\bar{f}$ amplitude (where $G$ is a KK graviton),
and  provide  a detailed
analysis of this new-physics effect in the light of 
our present knowledge on four-fermion processes at LEP and SLD 
\cite{ewg}.

In the next section,   we present
the analytical and numerical results for 
the angular distributions and total cross sections
for the process $e^+e^-\to f\bar{f}$ at the Z pole,
for both unpolarized and polarized initial beam and final states.
We also discuss possible effects on the model-independent
determination of the $Z$ mass through the $Z$ line shape.
Section 3 is devoted to our conclusions.

\section{Virtual-graviton effects on physical observables at the 
Z pole}

We consider the two-fermion production
in electron-positron collisions on the $Z$ pole
\beq
e^-(k_1) e^+(k_2) \to f(k_3) \bar{f}(k_4)
\eeq
where $k_i$ are the particle momenta. 

We want to compute the physical effects 
of the interference of the resonant $Z$ diagram with the 
amplitude where a virtual KK 
graviton is exchanged in the $s$-channel.
We define
the Mandelstam variables as $s=(k_1+k_2)^2 \,$ $[\sim M_Z^2]$, 
$t=(k_1-k_3)^2$, $u=(k_1-k_4)^2$, and  
neglect all the fermion masses.

The scattering amplitude in the momentum space of the graviton-mediated process
is given by a sum over all KK modes \cite{GRW}
\beq
{\cal A} =\frac{1}{\bar{M}_P^2}\sum_n \left\{ T_{\mu\nu}
\frac{P^{\mu\nu\alpha\beta}}{s-m^2_{G_n}}T_{\alpha\beta}
+\frac{1}{3}\left(
\frac{\delta-1}{\delta+2}\right)\frac{T^{\mu}_{\mu}
T^{\nu}_{\nu}}{s-m^2_{S_n}}\right\} \, ,
\label{amplitude}
\eeq
where $\bar{M}_P$ is the reduced Plank mass ($\bar{M}_P=M_P/\sqrt {8 \pi}$). 
The first term in Eq.(\ref{amplitude}) corresponds to graviton exchange,  
 the second to scalars exchange, with masses 
$m_{G_n}$ and $m_{S_n}$, respectively.
The $P^{\mu\nu\alpha\beta}$ is the projector of the 
graviton propagator, that in the unitary gauge is given by
\beq
P^{\mu\nu\alpha\beta}=\frac{1}{2}\left(\eta^{\mu\alpha}\eta^{\nu\beta}
+\eta^{\mu\beta}\eta^{\nu\alpha}\right)-
\frac{1}{3}\eta^{\mu\nu}\eta^{\alpha\beta}
+\dots
\eeq
where $\eta^{\mu\nu}$ is the Minkowski metric and dots 
represent the terms proportional to the momentum of the graviton propagator.
Since we are using the massless approximation for initial and final states,
 terms proportional to the 
trace of $T_{\mu\nu}$ vanish. 
Terms in the graviton propagator that are
proportional to the graviton momentum $q_{\mu}$ vanish, too, 
being $q^{\mu}T_{\mu\nu}=0$.
Hence, Eq.(\ref{amplitude}) becomes 
\beq
{\cal A}={\cal S}(s) 
T_{\mu\nu}T^{\mu\nu},~~~~
{\cal S}(s)=\frac{1}{\bar{M}_P^2} 
\sum_n \frac{1}{s-m^2_{G_n}}
\label{reducedA}
\eeq
The sum above can be easily computed in the continuum approximation for the
KK graviton spectrum \cite{GRW} 
\beq
{\cal S}(s)=\frac{1}{M_D^{2+\delta}} \int d^{\delta} q_T 
\frac{1}{s-q_T^2} = \frac{\pi^{\frac{\delta}{2}}}{M_D^4}
\Gamma(1-\frac{\delta}{2})
\left(-\frac{s}{M_D^2}\right)^{\frac{\delta}{2}-1}
\eeq
where we  used $m_{G_n}^2=q_T^2$, with $q_T$ the graviton momentum 
orthogonal to the brane.
In the interference with the resonant  
channel at the $Z$ pole, only  $ Im[{\cal S}(s)]$ will contribute,
with \cite{GRW}
\beq
Im [{\cal S}(s)]=-\frac{\pi}{M_D^{2+\delta}}\frac{S_{\delta-1}}{2}
s^{\frac{\delta-2}{2}}
\label{ImS}
\eeq
where, for $\delta=2n$,
$S_{\delta-1}=2\pi^n/(n-1)!$ and,  for $\delta=2n+1$,
$S_{\delta-1}=2\pi^n/\prod^{n-1}_{k=0} (k+\frac{1}{2})$, 
 with $n$ integer.
 While the real part of ${\cal S}$ (that is not relevant in our case !)
 is divergent and must be 
regularized by introducing an external cut off that parametrizes unknown 
new physics contributions in the ultraviolet region,
the imaginary part is finite and predictable,
only depending on the D-dimensional Plank scale $M_D$ and on
the number of extra dimensions $\delta$. Indeed,
the imaginary part  is finite 
because  it corresponds to the branch-cut singularity of real 
graviton emission. 

\

We now present how the SM cross sections and distributions
for the resonant process $e^+e^-\to f\bar{f}$
(assuming $f$  different from both $e^-$ and $\nu_e$) are modified 
after the inclusion
of the interference with the virtual-graviton exchange.
We will make a tree-level analysis, i.e. we will
neglect non-leading effects in the SM amplitude, such as 
the $s$-channel photon exchange, initial-state
radiation, and electroweak and strong corrections.
Although  such corrections will turn out to be in general larger than  the
virtual-graviton ones, we will assume that the main effect 
of the virtual-graviton amplitude arises 
from its interference with the tree-level
SM amplitude. We will also neglect terms corresponding to the 
squared amplitude of the graviton exchange.

One can find how the
SM differential distribution $d \sigma_{f}/d cos{\theta}$ for the process 
$e^+e^-\to f\bar{f}$ is modified by interference terms with the graviton
exchange, by means of Eq.(\ref{reducedA}) and 
Eq.(\ref{ImS}). 
At the $Z$ pole (i.e., for $s=M^2_Z$), one gets 
the new contribution
\bea
\frac{d \sigma^{\rm New}_f}{d \cos{\theta}}&=&
-\frac{N_cG_F S_{\delta-1}}{128 \sqrt{2}} 
\left(\frac{M_Z}{M_D}\right)^{2+\delta}\left(\frac{M_Z}{\Gamma_Z}\right)
\left\{g_A^eg_A^f\left(1-3\cos^2{\theta}\right)-
2 g_V^eg_V^f\cos^3{\theta}\right\}
\eea
with $\theta$  defined by
$t=-s(1-\cos{\theta})/2$. 
$G_F$ is the Fermi constant, $\Gamma_Z$ is the $Z$ total width, 
$g_V^f=T_3^f-2Q_f\sin^2{\theta_W}$ and $g_A^f=T_3^f$ are the standard
vectorial and axial $Z$ couplings (with $T_3^e=-1/2$),
and $N_c=3$  for quarks and 1 for leptons.

Correspondingly,
the two-fermion cross section at the $Z$ pole, is modified as follows
\beq
\frac{d \sigma_f}{d \cos{\theta}}=
\frac{9\pi}{2M_Z^2}\frac{\Gamma_{e^+e^-}\Gamma_{f\bar{f}}}{\Gamma_Z^2}
S_f(\cos{\theta})
\label{dsigma}
\eeq
where  
\beq
S_f(x)= 1+x^2-\Delta_1
\left(1-3x^2\right)
 +2A_eA_f\left(x+\Delta_2x^3\right) ,
\label{distt}
\eeq
and $\Gamma_{f\bar{f}}$ is the $Z$ partial width into $f\bar f$,
$\Gamma_{f\bar{f}}=\frac{N_c G_F M_Z^3}{6\pi\sqrt{2}}
\left((g_V^f)^2+(g_A^f)^2\right)$ ,
and $A_f=2 g_V^fg_A^f/\left((g_V^f)^2+(g_A^f)^2\right)$.
The graviton-interference effects are confined in the
coefficients  $\Delta_{1}$ and $\Delta_{2}$
that are $f$-flavour dependent
\footnote{A superscript $\Delta^f$ is understood wherever in the paper.}, with
\beq
\Delta_1=R_\delta~\frac{A_eA_f}{g_V^eg_V^f},~~~~~~~
\Delta_2=\frac{R_\delta}{g_A^eg_A^f}\, ,~~~
\label{delta12}
\eeq
and
\beq
R_\delta=\frac{\pi S_{\delta-1}}{32\sqrt{2}G_FM_Z^2}
\left(\frac{\Gamma_Z}{M_Z}\right)
\left(\frac{M_Z}{M_D}\right)^{2+\delta}.
\eeq

In table \ref{table1},
values of $R_\delta$ are shown for $\delta=2,3,4,5,6$, and
$M_D$ in the range (0.5,3.0) TeV.
The corresponding values of $\Delta_{1}$ and $\Delta_{2}$
for number of extra dimensions $\delta=2$,
in the three cases of final charged leptons, up-type quarks 
and down-type quarks,  
are given in table 
\ref{table2}. 

\noindent
Note that the whole dependence on the number of extra dimensions 
$\delta$ and
on the D-dimensional Plank scale $M_D$ is contained in $R_\delta$.
Values of $\Delta_{1}$ and $\Delta_{2}$ 
for a larger number of extra dimensions 
$\delta=3,4,5,6$ can be simply computed by multiplying the values 
in table \ref{table2} by the appropriate factors
$R_{3,4,5,6}/R_2$ obtained from table \ref{table1}.
\begin{table}
\begin{center}
\begin{tabular}{|r||c|c|c|c|c|}
\hline 
${\rm M_D}$ 
& $R_2$ & $ R_3$ & $ R_4$ & $R_5$ & $R_6$  \\ \hline \hline 
$0.5$ & $1.4\times 10^{-4}$ & $5.0\times 10^{-5}$ & $1.4\times 10^{-5}$ & $3.5\times 10^{-6}$ & $7.4\times 10^{-7}$ \\ \hline
$1$   & $8.5\times 10^{-6}$ & $1.6\times 10^{-6}$ & $2.2\times 10^{-7}$ & $2.7\times 10^{-8}$ & $2.9\times 10^{-9}$ \\ \hline
$2$   & $5.3\times 10^{-7}$ & $4.8\times 10^{-8}$ & $3.5\times 10^{-9}$ & $2.1\times 10^{-10}$ & $1.1\times 10^{-11}$ \\ \hline
$3$   & $1.1\times 10^{-7}$ & $6.4\times 10^{-9}$ & $3.0\times 10^{-10}$ & $1.2\times 10^{-11}$ & $4.4\times 10^{-13}$ \\ \hline
\end{tabular}
\caption[] {Numerical values of $R_{\delta}$
for $\delta=2,3,4,5,6$, and for a few  representative values of $M_D$ expressed
in TeV.}
\label{table1}
\end{center}
\end{table}
\begin{table}
\begin{center}
\begin{tabular}{|r||c|c|c|c|}
\hline
${\rm M_D}$
& $\Delta_1$ & $ \Delta_2$ & $ \Delta^{\prime}_1$ & 
$\Delta^{\prime}_2$ \\ \hline \hline 
 & ${\rm charged \, \, leptons}$ \\ \hline \hline 
$0.5$ & $2.2\times 10^{-3}$ & $5.4\times 10^{-4}$ & $1.1\times 10^{-3}$ 
& $1.1\times 10^{-3}$ \\ \hline
$1$ & $1.3\times 10^{-4}$ & $3.4\times 10^{-5}$ & $6.8\times 10^{-5}$ 
& $6.8\times 10^{-5}$ \\ \hline
$2$ & $8.4\times 10^{-6}$ & $2.1\times 10^{-6}$ & $4.2\times 10^{-6}$ 
& $4.2\times 10^{-6}$ \\ \hline
$3$ & $1.7\times 10^{-6}$ & $4.2\times 10^{-7}$ & $8.4\times 10^{-7}$ 
& $8.4\times 10^{-7}$ \\ \hline \hline
 & ${\rm u-type \, \,  \, quarks}$ \\ \hline \hline 
$0.5$ & $-1.9\times 10^{-3}$ & $-5.4\times 10^{-4}$ & $-9.5\times 10^{-4}$ & $-1.1\times 10^{-3}$ \\ \hline
$1$ & $-1.2\times 10^{-4}$ & $-3.4\times 10^{-5}$ & $-5.9\times 10^{-5}$ & $-6.8\times 10^{-5}$ \\ \hline
$2$ & $-7.4\times 10^{-6}$ & $-2.1\times 10^{-6}$ & $-3.7\times 10^{-6}$ & $-4.2\times 10^{-6}$ \\ \hline
$3$ & $-1.5\times 10^{-6}$ & $-4.2\times 10^{-7}$ & $-7.3\times 10^{-7}$ & $-8.4\times 10^{-7}$ \\ \hline \hline
 & ${\rm d-type \, \, \,quarks}$ \\ \hline \hline 
$0.5$ & $1.5\times 10^{-3}$ & $5.4\times 10^{-4}$ & $7.4\times 10^{-4}$ & $1.1\times 10^{-3}$ \\ \hline
$1$   & $9.1\times 10^{-5}$ & $3.4\times 10^{-5}$ & $4.6\times 10^{-5}$ & $6.8\times 10^{-5}$ \\ \hline
$2$   & $5.7\times 10^{-6}$ & $2.1\times 10^{-6}$ & $2.9\times 10^{-6}$ & $4.2\times 10^{-6}$ \\ \hline
$3$   & $1.1\times 10^{-6}$ & $4.2\times 10^{-7}$ & $5.7\times 10^{-7}$ & $8.4\times 10^{-7}$ \\ \hline
\end{tabular}
\caption[] {Numerical values of the coefficients $\Delta_{1,2}$, 
and $\Delta^{\prime}_{1,2} \,$, defined in 
eqs.(\ref{delta12}) and (\ref{ALRcos}),
respectively, for 
 final states of different flavors and $\delta=2$.
Some representative values of $M_D$ (expressed in TeV) are assumed.} 
\label{table2}
\end{center}
\end{table}

A first remarkable result in Eq.(\ref{dsigma}) is that the
contribution of the $Z$-graviton interference in the total cross
section at the $Z$ pole exactly vanishes, after integrating the
distributions $(1-3\cos^2{\theta})$ and $(\cos^3{\theta})$ in
Eq.(\ref{distt}) over the full range $(-1,1)$ of $\cos{\theta}$.
Indeed, this result holds in a larger class of processes.  We found
that the contribution to the {\it total } cross section of the
interference of the (non-necessarily resonant) amplitude for the
scattering $e^+e^-\to f\bar{f}$ mediated by spin-1 fields in the
$s$-channel (either axially or vectorially coupled) with the amplitude
mediated by a spin-2 field in the $s$-channel always vanishes.
\footnote{We checked that the results on fermion scattering mediated
by virtual graviton exchanges in \cite{GRW} satisfy this property.}
Therefore, in the total cross section for the scattering $e^+e^-\to
f\bar{f}$, with $f\neq e,\nu_e$, virtual graviton effects are
suppressed, being due, at leading order, to the squared amplitude of
the graviton-exchange channel.

On the other hand, the non-trivial effect of the graviton interference
on angular distributions gives rise to cross-section variations with
respect to the SM values, whenever non-uniform angular conditions are
applied.  For instance, a forward-backward symmetric angular cut along
the beams gives rise to a $\Delta_1$ dependence of the cross section
on the $Z$ resonance.  An asymmetric cut produces a $\Delta_2$
dependence, too.

Accordingly, we present now how the different asymmetries determined
at LEP and SLD \cite{ewg} are modified by the virtual-graviton
interference at the $Z$ pole.  In particular, we first show how the SM
predictions are modified for the forward-backward asymmetries
$A^f_{FB}$.  A non-vanishing effect is also found and shown in the
cases of the left-right polarization asymmetries $A^f_{LR}$, the
left-right forward-backward asymmetries $A^{FB}_{LR}(f)$, and the
$\tau$ polarization asymmetries $P_{\tau}(\cos{\theta})$.

For the forward-backward asymmetry $A^f_{FB}$, defined as
\beq A^f_{FB}=
\frac{1}{\sigma_f}
\left[
\int_{0}^1
\left(\frac{d \sigma_f}{d \cos{\theta}}\right) d \cos{\theta}
-\int_{-1}^0 
\left(\frac{d \sigma_f}{d \cos{\theta}}\right) d \cos{\theta}\right]~~~,
\label{asymFB}
\eeq
we obtain by Eq.(\ref{dsigma})
\beq
A^f_{FB}=
\frac{3}{4} A_eA_f\left(1+\frac{\Delta_2}{2}\right) \, .
\label{totalAFB}
\eeq
Since $\Delta_1$ in Eq.(\ref{distt}) multiplies a $\cos{\theta}$-even
term, the $\Delta_1$ dependence drops in the $A^f_{FB}$
definition. The numerical deviation from the $A^f_{FB}$ standard-model
value can be obtained by Eq.(\ref{totalAFB}) from the $\Delta_2$
values shown in table \ref{table2}. For $\delta \geq 2$ and $M_D>1$
TeV, one gets $|\Delta_2|<10^{-4}$ for any flavour $f$.  Hence, the
virtual-graviton interference gives rise to a deviation in $A^f_{FB}$
that is much smaller than its present experimental error (about 1\%
\cite{ewg}) from the LEP experiments.

One could think about a new kind of asymmetry that maximizes the
sensitivity to the graviton-$Z$ interference.  In Eq.(\ref{distt}),
the SM distribution in $\cos{\theta}$ has an even part
[$(1+\cos^2{\theta})$] and an odd part [$\cos{\theta}$] that behave
differently from the graviton-interference even part
[$1-3\cos^2{\theta}$] and odd part [$\cos^3{\theta}$], respectively.
Then, it is possible to define a new angular asymmetry $A_G^f$ which
vanishes at the tree-level in SM,
\bea
A_G^f&=&\frac{1}{\sigma_f}\left[
\int_{-c^{\star}}^0
\left(\frac{d \sigma_f}{d \cos{\theta}}\right) d \cos{\theta}
-
\int_{-1}^{-c^{\star}}
\left(\frac{d \sigma_f}{d \cos{\theta}}\right) d \cos{\theta}
\right]
\nonumber \\
&+& \frac{1}{\sigma_f}\left[ \int_{0}^{c^{\star}}
\left(\frac{d \sigma_f}{d \cos{\theta}}\right) d \cos{\theta}
-\int_{c^{\star}}^1
\left(\frac{d \sigma_f}{d \cos{\theta}}\right) d \cos{\theta}
\right]~~~ ,
\label{asym}
\eea
where $c^{\star}=-(1+\sqrt{2})^{-1/3}+(1+\sqrt{2})^{1/3}$ 
is the (unique) real solution of the equation
\beq
\int_0^{c^{\star}}(1+x^2)d x-\int_{c^{\star}}^1(1+x^2)d x=0,
\eeq
corresponding to $\theta^{\star}=\arccos c^{\star} \simeq 53^0$.
From Eq.(\ref{dsigma}), one then gets for $A_G^f$ a maximal sensitivity
to graviton effects
\beq
A_G^f\simeq -0.5764\, \,  \Delta_1 \, .
\label{AG}
\eeq
Note that, although the $A_G^f$ angular asymmetry shows maximal sensitivity
to the graviton interference at tree level,
the inclusion of angular-dependent electroweak radiative corrections can 
(at least partly) spoil the goodness of the definition in Eq.(\ref{asym}).
We do not go further into this point, and leave it as a suggestion
for new analysis of experimental data.

We are now going to consider the cross-section at the $Z$ pole, 
for a longitudinally polarized electron beam. 
Eq.(\ref{dsigma}) can be easily generalized 
to the case of an electron beam with polarization
$\psi_p=P_e\psi$ (with the projector $P_e=(1+p_e\gamma_5)/2$) as follows
\bea
\frac{d \sigma_f(p_e)}{d \cos{\theta}}&=&
\frac{9\pi}{2M_Z^2}\frac{\Gamma_{e^+e^-}\Gamma_{f\bar{f}}}{\Gamma_Z^2}
\hat{S_f}(\cos{\theta},p_e)
\label{polarized} \\
\hat{S_f}(x,p_e)&=&
\left(1-p_eA_e\right)
\left(1+x^2-\hat \Delta_1(p_e)\left(1-3x^2\right)\right)\nonumber \\
 &+&2A_f\left(A_e-p_e\right)
\left(x+\hat \Delta_2(p_e)x^3\right)
\nonumber
\eea
where the coefficients 
$\hat{\Delta}_1(p_e)$, and $\hat{\Delta}_2(p_e)$, 
 are given by
\bea
\hat \Delta_1(p_e) &=& R_\delta
\frac{A_eA_f}{\left(1-p_eA_e\right)}
\left(\frac{1}{g_V^eg_V^f}-\frac{p_e}{g_A^eg_V^f}
\right)
,\nonumber \\
\hat \Delta_2(p_e) &=& R_\delta
\frac{A_e}{
\left(A_e-p_e\right)}\left(\frac{1}{g_A^eg_A^f}-\frac{p_e}{g_V^eg_A^f}\right)\,.
\eea
Note that $\hat{\Delta}_{1,2}(p_e=0) =\Delta_{1,2}$, and 
$\hat{S_f}(x,0)=S_f(x)$.
For instance, for $M_D= 1$ TeV, $\delta = 2$, and charged-lepton final states,
 one has, for $p_e=1$, 
 ${\hat \Delta}_1 = 1.5 \times 10^{-4}$ and
 ${\hat \Delta}_2 = 7.4 \times 10^{-5}$,
 while, for $p_e=-1$, 
 ${\hat \Delta}_1 = 1.2 \times 10^{-4}$ and
 ${\hat \Delta}_2 = 6.3 \times 10^{-5}$.

From Eq.(\ref{polarized}), using the polarized {\it total} cross
section $\sigma_f(p_e)$
(that, as in the $\sigma_f$ case, is graviton-interference independent), 
one can determine
the differential left-right asymmetry, defined as
\beq
\frac{dA_{LR}^f(p_e)}{d \cos{\theta}}= \frac{1}{\sigma_f(p_e)+\sigma_f(-p_e)}
\left( \frac{d \sigma_f(-p_e)}{d\cos{\theta}}
-\frac{d \sigma_f(p_e)}{d\cos{\theta}}\right) \, .
\label{asympol_def}
\eeq
The result is
\beq
\frac{dA_{LR}^f(p_e)}{d \cos{\theta}}=
\frac{3}{8}p_e\left\{A_e\left(1+\cos^2{\theta}-
\Delta^{\prime}_1\left(1-3\cos^2{\theta}\right)\right)
+2A_f\left(\cos{\theta}+\Delta^{\prime}_2\cos^3{\theta}\right)\right\} \, ,
\label{asympol_res}
\eeq
where $\Delta_{1}^{\prime}$ and $\Delta_{2}^{\prime}$ are given by
\beq 
\Delta_1^{\prime}=R_{\delta} \left(\frac{A_f}{g_A^eg_V^f}\right)
=\left(\frac{g_V^e}{g_A^eA_e}\right)\Delta_1,~~~~~~~~
\Delta_2^{\prime}=R_{\delta} \left(\frac{A_e}{g_V^e g_A^f}\right)
=\left(\frac{g_A^eA_e}{g_V^e}\right)\Delta_2\, .
\label{ALRcos}
\eeq

Numerical values of $\Delta_{1}^{\prime}$ and $\Delta_{2}^{\prime}$
for $\delta = 2$, and different final
states, are reported in table \ref{table2}. One gets, for $\delta \geq
2$ and $M_D>1$ TeV, $|\Delta_{1,2}^{\prime}|<10^{-4}$ for any flavour
$f$.  In the total left-right asymmetry, the $Z$-graviton contribution
vanishes, and one recovers the SM tree-level value
$A_{LR}^f(p_e)=p_eA_e$.  As discussed previously, the pattern of
electroweak radiative corrections could alter these leading-order
results in a non-trivial way.

 The left-right forward-backward asymmetry,  
defined as 
\beq
A_{LR}^{FB}(f)= 
\int_{0}^1
\left(\frac{d A_{LR}^f(p_e)}{d \cos{\theta}}\right) d \cos{\theta}
-\int_{-1}^0 
\left(\frac{d A_{LR}^f(p_e)}{d \cos{\theta}}\right) d \cos{\theta}\, ,
\eeq
has been also measured at SLC.
For this observable, we obtain the following modification of the SM value
\beq
A_{LR}^{FB}(f)=\frac{3}{4}p_eA_f\left(1+\frac{\Delta_2^{\prime}}{2}\right) \, .
\eeq
Also in this case, the graviton-interference effects are much below
the experimental errors \cite{ewg}.  Here, too, one could try to
introduce ``most sensitive" variables of the kind of $A_G^f$ suggested
in Eq.(\ref{asym}), hence getting a new asymmetry proportional to
$\Delta_1^{\prime}$.

We complete now the list of the asymmetries measured at LEP/SLC,
by  considering the $\tau$ polarization asymmetry, 
that is defined as
\beq
P_{\tau}(\cos{\theta})=\frac{d\sigma^R_{\tau}(\cos{\theta})-
d\sigma^L_{\tau}(\cos{\theta})}{
d\sigma^R_{\tau}(\cos{\theta})+d\sigma^L_{\tau}(\cos{\theta})}
\eeq
where $d\sigma^R_{\tau}(\cos{\theta})$ and 
$d\sigma^L_{\tau}(\cos{\theta})$ stand for the  differential 
cross sections $\frac{d\sigma_{\tau}}{d\cos{\theta}}(p_{\tau})$ 
for the production of a right-handed ($p_{\tau}=1$) and 
a left-handed ($p_{\tau}=-1$) $\tau$-lepton, respectively.
The effect of the graviton interference at the $Z$ pole gives in this case
\beq
P_{\tau}(\cos{\theta})=-\frac{A_{\tau}\left(1+\cos^2{\theta}-
\Delta_1^{\prime}\left(1-3\cos^2{\theta}\right)\right)+2A_e\left(\cos{\theta}+
\Delta^{\prime}_2\cos^3{\theta}\right)}
{1+\cos^2{\theta}-\Delta_1\left(1-3\cos^2{\theta}\right)
+2A_eA_{\tau}\left(\cos{\theta}+\Delta_2\cos^3{\theta}\right)}
\eeq
Here, too,  the virtual graviton effect is far from being detectable with the
present experimental precision \cite{ewg}.
 
\

\ 


Finally, we want to discuss the possibility how the  $Z$-graviton
interference affects the  model-independent determination of $M_Z$
 from the $Z$ line shape \cite{MI}.  We showed that the
{\it total} cross section on the $Z$ resonance is not altered at tree
level by $Z$-graviton interference effects.  On the other hand, any
cross section measurement is affected by angular cuts that will give
rise to finite effects.

The numerical value of the $Z$ mass is closely related to the peak
position of the  $Z$ line shape.
In the resonant region around the $Z$ pole, 
the differential cross section is given by
\bea
\frac{d \sigma_f}{d \cos{\theta}}&=&
\left(\frac{9\pi}{2M_Z^2}\frac{\Gamma_{e^+e^-}\Gamma_{f\bar{f}}}{\Gamma_Z^2}
\right)
F(s)\left\{1+\cos^2{\theta}+2A_eA_f\cos{\theta}
\right. \nonumber \\
&-&\left. G(s)\left[\Delta_1\left(1-3\cos^2{\theta}^2\right)
 -2A_eA_f\Delta_2\cos^3{\theta}\right]\right\} \, ,
\label{dsigma_res}
\eea
with $\Delta_1$ and $\Delta_2$ defined in Eq.(\ref{delta12}).
The function $F(s)$ is the usual Breit-Wigner resonance shape
that is used at LEP to fit the $Z$ mass value \cite{PDG},
while  $G(s)$ arises from the $Z$-graviton interference,
and can modify the Breit-Wigner shape 
\beq
F(s)=\frac{s\Gamma_Z^2}{\left(s-M_Z^2\right)^2+M_Z^2\Gamma_Z^2}
,~~~~G(s)=\left(\frac{s}{M_Z^2}\right)^{\frac{\delta}{2}} \, .
\eeq
We are  neglecting the contribution of the real
part of the graviton mediated amplitude, 
that is suppressed with respect
to its imaginary part by terms of order $(s-M_Z^2)/M_Z^2$.
After integrating over the polar angle the differential cross section in 
Eq.(\ref{dsigma_res}) over the interval $(-c^{\prime},c^{\prime})\,$
[with $0<c^{\prime}<1$], 
one gets
\bea
\sigma^{c^{\prime}}_f(s)&=&\left(
\frac{9\pi}{2M_Z^2}\frac{\Gamma_{e^+e^-}\Gamma_{f\bar{f}}}{\Gamma_Z^2}\right)
F(s)\left\{I_1-I_2\Delta_1 G(s)\right\}
\nonumber \\
I_1&=&\int_{-c^{\prime}}^{c^{\prime}}d x \left(1+x^2\right),~~~~
I_2=\int_{-c^{\prime}}^{c^{\prime}}d x \left(1-3x^2\right) \, .
\label{sigma_res}
\eea
The effect of the $Z$-graviton interference arises when
considering angular cuts $c^{\prime}<1$, while, for $c^{\prime}=1$,
one has $I_2=0$, and no effect  on the Breit-Wigner shape is observed.

 Then, in order to estimate the shift of the value of fitted $Z$ mass
 due to graviton-interference, one has to determine how much the
 position of the maximum of $\sigma^{c^{\prime}}_f(s)$ is modified by
 the function $G(s)$.  To do that, one has to solve the equation
 $\frac{d}{d s}\sigma^{c^{\prime}}_f(s)=0$. By Eq.(\ref{sigma_res}),
 this is equivalent to solve the equation
\beq
F^{\prime}(s)\left(1-\frac{\Delta_1 I_2}{I_1}G(s)\right)-
F(s)G^{\prime}(s)\frac{\Delta_1 I_2}{I_1}=0 \, ,
\label{maximum}
\eeq
where the primed indices indicate the first derivative with respect to $s$.
 To find the solution of Eq.(\ref{maximum}), 
it is convenient to use 
perturbation theory in the parameter $\Delta_1$.
At the first order in $\Delta_1$, we make the ansatz
that the maximum of the graviton-modified distribution
is at the $Z$ mass value
\beq 
\bar{M}_Z^2=\hat{M}_Z^2+\delta_{M_Z^2} \, ,
\label{solu}
\eeq
where $\hat{M}_Z^2$ is the peak position  of the 
Breit-Wigner distribution $F(s)$
[that is $F^{\prime}(\hat{M}_Z^2)=0$  and 
$\hat{M}_Z^2=M_Z^2\sqrt{1+\Gamma_Z^2/M_Z^2}$], and $\delta_{M_Z^2}$  
is of order of ${\cal O}(\Delta_1)$.
By substituting $s=\bar{M}^2_Z$ in Eq.(\ref{maximum}),  making 
an expansion in $\delta_{M_Z^2}$,  and retaining only the first-order
terms, one has,
\beq
\delta_{M_Z^2}=-\frac{\Gamma_Z^2 \delta\Delta_1}{4}\left(\frac{I_2}{I_1}\right)
\left(1
+{\cal O}(\frac{\Gamma^2_Z}{M^2_Z})\right) \, .
\label{shift2}
\eeq
The shift in the $Z$ peak turns out to be quite small, since, it is
not proportional to the squared $Z$ mass, but to the squared $Z$
width.  Keeping only the leading term in the $\Gamma^2_Z / M^2_Z$
expansion, we find that $M_Z$ is shifted by the quantity
\beq
\Delta M_Z=-\frac{\Gamma_Z^2 \Delta_1\delta }{8M_Z}\left(\frac{I_2}{I_1}
\right).
\label{shift1}
\eeq
Of course the shift is critically dependent on the angular cuts entering
into the integrals $I_{1,2}$.
For $c^{\prime}=1/\sqrt{3}$,
 $\delta=2$, $M_D=1$ TeV, and charged-lepton final states,
the $Z$-graviton interference  shifts the $Z$
resonance peak by about --1.4 keV.
For up-type and down-type quarks, the shift is about 1.2 keV and --1.0
keV, respectively.
The different sign for different final states would partly cancel
the effect, that is anyhow much smaller than the present 
$M_Z$ experimental error (2.1 MeV \cite{PDG}).

\section{Conclusions}

In scenarios with quantum gravity propagating in large extra
dimensions, we computed the effects of the interference of the virtual
KK graviton-exchange amplitude with the resonant amplitude on the $Z$
pole, on $e^+e^-\to f \bar f$ physical observables.  Remarkably,
results are finite and predictable in terms of the fundamental
$D$-dimensional Plank scale $M_D$.  The corresponding effect on total
cross sections at tree-level vanishes, and this result remains valid
for $e^+e^-\to f \bar f$ (with $f\neq e, \nu_e$) far from the $Z$
resonance.  On the other hand, non-trivial modifications of the
angular distributions give rise to finite effects whenever angular
cuts are applied, and $Z$-graviton interferences contribute to the
forward-backward asymmetries.  We also computed initial and final
polarization asymmetries.  In general, the {\it relative} magnitude of
the $Z$-graviton interference effects on all these observables is of
the order $10^{-4}$ for number of extra dimensions equal to  2 and
$M_D\sim 1$ TeV, that is 2 orders of magnitude smaller than the
present experimental errors.  Even a future linear collider operating
in the {\it GigaZ} mode
\cite{Aguilar-Saavedra:2001rg}
(that is expected to improve the errors on many $Z$-pole precision
measurements by about an order of magnitude) seems not to be 
sensitive enough to detect such effects. 

\noindent
Finally, we discussed the shift in  $M_Z$ from the
model-independent determination due to the modification of the
Breit-Wigner resonant shape.  Shifts in $M_Z$ are found at most of
the order 1 keV.  In conclusion, although appealing on theoretical
grounds, virtual graviton exchanges do not presently affect
$Z$-resonance observables at a detectable level.

\section*{Acknowledgments}
We would like to thank Gian Giudice  and Riccardo Rattazzi
for useful discussions. E.G. and A.D. would also like to thank the Physics
Department of University of Roma ``La Sapienza''
for its kind of hospitality during the preparation of this work.
A.D. also thanks Academy of Finland (project number 48787) 
for financial support.
\vskip 2cm

\end{document}